\documentclass[onecolumn,showpacs,showkeys,preprintnumbers,amsmath,amssymb]{revtex4}

\usepackage{graphicx}% Include figure files
\usepackage{dcolumn}% Align table columns on decimal point
\usepackage{bm}% bold math

\begin{document}

%\preprint{}
\title{Direct manifestation of Ehrenfest's theorem in the infinite square well model}

\author{Chyi-Lung Lin}
\email{cllin@scu.edu.tw}
%\email[]
%\thanks{}
\affiliation{Department of Physics,\\ Soochow University, \\Taipei,Taiwan, R.O.C.}

%\author{\thanks{ E-mail: \email{ }}}

%\institute{Department of Physics,\\ Soochow University, \\Taipei,Taiwan, %R.O.C.}

%\pacs{03.65.w}{ Quantum mechanics}


%\date{\today}

\begin{abstract}
{ 

Ehrenfest's theorem in the infinite square well is up to now only manifested indirectly. The manifestation of this theorem is first done in the  finite square well, and then consider the infinite square well as the limit of the finite well. 
For a direct manifestation, we need a more precise formula to describe the degree of infiniteness of the divergent potential energy. We show that the potential energy term term,  $V(x) \Psi_n(x)$, which is the product of the potential energy and the energy eigenfunction, is a well defined function which can be expressed in terms of Dirac delta functions. 
This means that the infinity in this model is not that vague but has obtained a specification.
This results that expectation values can be calculated precisely and  Ehrenfest's thereom can be confirmed directly.

}

\end{abstract}

%\keywords{}



%----------------------------------

\maketitle


\section{\label{sec:1}  Introduction}

The infinite square well is a model using infinitely high potential barrier to confine particles inside a well. This is a basic model in quantum mechanics, and is a standard model for confining particles. 
We also have other models using infinitely large potential energy, such as the Dirac delta function potential. Belloni and Robinett have given a review on these models \cite{b.r1}. 

Curiously, for infinite square well, we still do not have a direct manifestation of Ehrenfest's theorem in this fundamental model. In contrast to the Dirac delta function potential, the infinity in the infinite square well has not been clearly described. 
Simply using the notation as $\infty$ is vague, we have no guide on how to handle this infinity.
We need a more precise formula to describe the degree of infiniteness of the divergent potential energy, so that we can perform precise calculation for expectation values. 
%It lacks a formula to specify the infinity. We thus have no guide on how to handle this infinity. This results that expectation values can not be calculated and Ehrenfest's theorem can not be confirmed. 
%This is the main reason why so far there does not exist a direct manifestation of Ehrenfest theorem in this model.
In this paper, we investigate this problem and show that Ehrenfest theorem can directly be confirmed.

%
%For the delta function we have a specification formula to specify the infinity, that is with the total area of unity. 
%In contrast to the Dirac delta function potential, the infinity in the infinite square well has not been clearly described. It lacks a specification formula like the one for the delta function. We thus have no guide on how to handle this infinity. This is the reason why so far there is not a direct manifestation of Ehrenfest theorem in this model. 
%In this paper, we investigate this problem and show that Ehrenfest theorem can directly be confirmed.

%
%
The potential energy $V(x)$ of the infinite square well is described by
\begin{equation}
\label{Eq.1A-1}
V(x)=\left\{
\begin{array}{ll}
0,&        0<x<L, \\
 \infty,&      \text{otherwise}. 
\end{array}
\right.
\end{equation}
The time independent Schr\"odinger equation  is
\begin{equation}
\label{Eq.1A-2}
-\frac{\hbar^2}{2m} \Psi''(x)+V(x) \Psi(x)=E \Psi(x).
\end{equation} 
For $V(x)$ defined in  Eq.~(\ref{Eq.1A-1}), 
the energy eigenfunctions $\Psi_n (x)$ and the eigenvalues $E_n$ of Eq.~(\ref{Eq.1A-2}) are well-known \cite{b.r2,b.r3,b.r4,b.r5, b.r6,b.r7,b.r8}. We have
\begin{eqnarray}
\label{Eq.1A-3}
&& \Psi_n (x)=\left\{
\begin{array}{ll}
\sqrt\frac{2}{L}\ \sin(k_n  x),&        0<x<L, \\
  0,&      \text{otherwise}.
\end{array}
\right.
\\
&& E_n=\hbar^2  k_n^2/ 2 m,
\label{Eq.1A-4}
\end{eqnarray}
where $k_n= n \pi/L$, $n=1,2,3...$.  
%%%%
%%%%%

Though we have exact solutions and these solutions are simple; however, the direct manifestation of Ehrenfest's theorem in this model is so far a lack. The reason is that we can not calculate the expectation value of the force operator, which corresponds to the term $-dV(\hat{x})/d \hat{x}$.  We note that the force is infinitely large at the two sides of the well, while the probability density $\Psi^{*}(x) \Psi(x)$ is zero there.  Hence, calculating the expectation value of the force operator, we encounter the ambiguity as the product of infinity and zero \citep{b.r9}.

To solve this ambiguity, one may consider the infinite square well as a limiting case of a finite well  \citep{b.r9, b.r10}. 
Rokhsar calculated the expectation value of the force in a finite well, and then took the limit of the height of the potential energy $V_0 \to \infty$ to confirm Ehrenfest theorem in the infinite square well \citep{b.r9}.

We note that the difficulty for a direct manifestation lies in lacking informations about the specification of the infinity used in the potential energy.
For the case of Dirac delta function $\delta(x)$, whcih is usually defined as
\begin{equation}
\label{Eq.1A-5}
\delta(x)=\left\{
\begin{array}{ll}
0,&       x \neq 0, \\
 \infty,&      x=0. 
\end{array}
\right.
\end{equation}
The infinity in the delta function is specified by the formula
\begin{equation}
\label{Eq.1A-6}
\int_{-\infty}^\infty \delta(x) dx = 1.
\end{equation}
Eq.~(\ref{Eq.1A-6}) is the specification formula for the delta function. This formula shows how to handle the infinity in the delta function. 
%We may consider the delta function as the limit of the finite square step function, defined as: 
%\begin{equation}
%\label{Eq.1A-7}
%f(x, \epsilon)=\left\{
%\begin{array}{ll}
%\frac{1}{\epsilon},&   \frac{-
%\epsilon}{2}< x <\frac{\epsilon}{2}, 
%\\
% 0,&      \text{otherwise}. 
%\end{array}
%\right.
%\end{equation}
%
%Simply taking the limit $\epsilon \to 0$ leads directly to Eqs.~(\ref{Eq.1A-5}-\ref{Eq.1A-6}). But, on the hand, we do have a functional form to this limit, that is 
%
%
%\begin{equation}
%\delta(x)= \frac{1}{2 \pi} \int_{-
%%%\infty}^\infty e^{- i x y} dy.
%\label{Eq.1A-8}
%\end{equation} 
%
%This is a very useful formula for describing the delta function. 
In the same spirit, we seek a specification formula for the $V(x)$ in  the infinite square well model.
We discuss this in Sec. II. We show the direct manifestation of Ehrenfest theorem  in Sec. III, and we discuss more about the function form of $V(x) \Psi(x)$ in Section IV.
%
%The potential energy of a finite well may also be described as: $V(x)=V_0 [\theta(-x)+ \theta(x-L)]$. Taking the limit $V_0 \to \infty$ simply leads to the $V(x)$ in Eq.~(\ref{Eq.1A-1}). 
%This is not the functional form we are looking for.Instead of this way, we look for an approach to reach the limit of a finite square well physically.

%
%

\section{\label{sec:2} The derivation of the functional form of $V(x) \Psi_n(x)$}

To explore such a  specification formula of $V(x)$, we work in an opposite way. We start from wave functions $\Psi_n(x)$ and substitute these wave functions into the Schr\"odinger equation, we can then determine the corresponding \textit{functional} form of the potential energy term.

For the infinite square well, the wave function in Eq.~(\ref{Eq.1A-3}) can be rewritten in a more compact form as 
\begin{eqnarray}
\label{Eq.2A-1}
&& \Psi_n(x)= u_n(x) \theta(x)\theta(L-x).\\
\label{Eq.2A-2}
&& u_n(x)= \sqrt \frac{2}{L} \sin(k_n  x).
\end{eqnarray}
As we are to derive the form of $V(x)$ from the wave function,  we then rewrite the Schr\"odinger equation as:
\begin{equation}
\label{Eq.2A-3}
V(x) \Psi(x) = \frac{\hbar^2}{2m} \Psi''(x)+E \Psi(x).
\end{equation} 
For $\Psi(x)$ of the form as that in Eq.~(\ref{Eq.2A-1}), the $\Psi''(x)$ can be calculated by the following formula
\begin{eqnarray}
&& \Psi(x)=f(x) \theta(x) \theta(L-x)
\nonumber\\
&&\Psi''(x)=f''(x) \theta(x) \theta(L-x) + f'(x) \delta(x)  - f'(x)  \delta(L-x) .
\label{Eq.2A-4}
\end{eqnarray}
Substituting Eq.~(\ref{Eq.2A-1}) into Eq.~(\ref{Eq.2A-3}) then yields

\begin{equation}
V(x) \Psi_n (x)
=
 \frac{\hbar^2}{2m}   \Big[  \delta(x) -   \delta(L-x) \Big] U_n'(x).
\label{Eq.2A-5}
\end{equation}
Substituting Eq.~(\ref{Eq.2A-2}) into Eq.~(\ref{Eq.2A-5}), we have 
\begin{equation}
V(x) \Psi_n (x)
=\sqrt {\frac{2}{L}} \frac{\hbar^2}{2m}  k_n \Big[  \delta(x) -  \cos(k_n L)  \delta(L-x) \Big].
\label{Eq.2A-6}
\end{equation}
We have then obtained the functional form of $V(x) \Psi_n(x)$. It had already been argued  that the potential energy term, $V(x) \Psi_n(x)$,  should contain a delta function at each side of the well \cite {b.r10, b.r11}. Eq.~(\ref{Eq.2A-6}) shows this result explicitly. 

%Eq.~(\ref{Eq.2A-6}) shows the precise functional form of the $V(x) \Psi_n(x)$ in the infinite square well.
Eq.~(\ref{Eq.2A-6}) is an amazing formula. If we look back into the definition of $V(x)$ in Eq.~(\ref{Eq.1A-1}), then we see that the infinity in the $V(x)$ is not that unmanageable.
That is, although $V(x)$ is a divergent quantity outside the well, however $V(x) \Psi_n(x)$ is indeed a well-defined function which is described in terms of Dirac delta functions, as shown in the right side of   Eq.~(\ref{Eq.2A-6}). 
%We may view Eq.~(\ref{Eq.2A-6}) as the specification formula for the divergent $V(x)$. 
This result should be a stepping stone for further  investigations on the infinite square well model when there are ambiguities in calculating expectation values.
One of the application of this result is that we can directly confirm Ehrenfest's theorem.

%This is a step better than using the notation $\infty$. 

%Though the infinity used in the V(x) of  Eq.~(\ref{Eq.1A-1}) is not well specified, however, from the $\Psi_n(x)$, the notation of $\infty$ in $V(x)$ is now having a specification; a step better than using the notation $\infty$. 
%If we are to confine particles inside a well with $\Psi_n(x)$ as the energy eigenstates, then the potential energy $V(x)$ should have such a property described in Eq.~(\ref{Eq.2A-6}).
%We may analyze the term $V(x) \Psi_n(x)$ in the left side of Eq.~(\ref{Eq.2A-6}).
%
%We  note that $\Psi_n(x)$ defined in Eq.~(\ref{Eq.2A-1}) contains the factor $\theta(x) \theta(L-x)$. Thus, for the term  $V(x) \Psi_n(x)$ in Eq.~(\ref{Eq.2A-6}), there is the factor $\theta(x) \theta(L-x)$ accompanied with the $V(x)$.
%
%Because the divergence nature in the $V(x)$ of the infinite square well, the accompany of this factor with $V(x)$ shows how to handle the infinity in the $V(x)$. 
%The accompany of this factor with $V(x)$ shows how to handle the infinity in the $V(x)$. 
%That is, though $V(x)$ is a divergent quantity outside the well, however $V(x) \theta(x) \theta(L-x)$ is a well behaved function, and $V(x) \Psi_n(x)= V(x) \theta(x) \theta(L-x) u_n(x)$ is indeed a well-defined function which is described in the right side of   Eq.~(\ref{Eq.2A-6}).

%
%
\section{\label{sec:3} The verification of Ehrenfest's theorem in the infinite square well }

We now show that Ehrenfest's theorem for time-evolved wave packets in the infinite square well can be manifested.
The time evolution of a general wave packet $\Psi(x,t)$ is as follows
\begin{eqnarray}
\label{Eq.3A-1}
\Psi(x,t)=\sum_{n=1}^{\infty} a_n \Psi_n(x) e^{- i  \omega_n t}. 
\end{eqnarray}
where $\omega_n=E_n/\hbar$, and $\sum_{n=1}^{\infty} \mid a_n \mid^2=1$. 
To verify Ehrenfest's theorem, we need to verify the following formula:
\begin{equation}
\label{Eq.3A-2}
\frac{d}{dt} \langle\Psi(t) \mid \hat{p} \mid \Psi(t)\rangle  =  -\langle\Psi(t) \mid \frac{dV(\hat{x})}{d\hat{x}} \mid \Psi(t)\rangle.
\end{equation}
The calculation of the right side of  Eq.~(\ref{Eq.3A-2}) is related to the calculation of $\Psi_n(x)  (dV(x)/dx) \Psi_j(x)$. We note that
\begin{eqnarray}
&& \Psi_n(x)  \frac{dV(x)}{dx} \Psi_j(x)
\nonumber\\
&& = \frac{d}{dx}  [\Psi_n(x) V(x) \Psi_j(x)]-
\frac{d \Psi_n(x)}{dx}   [V(x) \Psi_j(x)]-
[\Psi_n(x) V(x)] \frac{d \Psi_j(x)}{dx}.
\label{Eq.3A-3}
\end{eqnarray}
For the right side of Eq.~(\ref{Eq.3A-3}), the first term makes no contribution when taking the integration. This is because after the integration of the first term over the range of  $x$ , $[0, L]$, the result is the value of the boundary term $\Psi_n(x) V(x) \Psi_j(x) $ calculated at the two sides of the well. And this yields zero by using Eq.~(\ref{Eq.2A-6}) and Eqs.~(\ref{Eq.2A-1}-\ref{Eq.2A-2}).
Using again Eq.~(\ref{Eq.2A-6}), the other two terms can be calculated, and we obtain
\begin{equation}
\frac{d \Psi_n(x)}{dx} V(x)\Psi_j(x)
=
\frac{\hbar^2}{m L}\ k_n k_j\ \theta(x) \theta(L-x)\ [\delta(x)-(-1)^{n+j} \delta(L-x)]. 
\label{Eq.3A-4}
\end{equation}
And then we have
\begin{eqnarray}
&&\int_{-\infty}^{\infty} \frac{d \Psi_n(x)}{dx} V(x)\Psi_j(x) dx
\nonumber\\
&&=\frac{\hbar^2}{m L}\ k_n k_j \int_{0}^{L} [\delta(x)-(-1)^{n+j} \delta(L-x)] dx
\nonumber\\
&& =\frac{\hbar^2}{2 m L}\ k_n k_j\ \beta_{n j}.
\label{Eq.3A-5}
\end{eqnarray}

where $\beta_{n j}=1-(-1)^{n+j} $. Above, we have used the following results

\begin{eqnarray}
&&\int_{0}^{L} \delta(x)\ dx = \frac{1}{2},
\label{Eq.3A-6}
\\
&&\int_{0}^{L} \delta(L-x)\ dx = \frac{1}{2}.
\label{Eq.3A-7}
\end{eqnarray}
These results are due to the even function property of the delta function.
%so that the integration of the delta function, over the range  of $x$, $[0, L]$, is equal to one half of that over the range, $[-L, L]$.
%
%
We then have
\begin{equation}
\label{Eq.3A-8}
\langle\Psi_n(x) \mid \frac{dV(\hat{x})}{d\hat{x}} \mid \Psi_j(x)\rangle
=
-\frac{\hbar^2}{ m L}\ k_n k_j\ \beta_{n j}.
\end{equation}
And then we obtain the final result
\begin{eqnarray}
\label{Eq.3A-9}
&&\langle\Psi(t) \mid \frac{dV(\hat{x})}{d\hat{x}} \mid \Psi(t)\rangle 
\nonumber\\
 &=&\int_{-\infty}^{\infty} \Psi^*(x,t) \frac{dV(x)}{dx} \Psi(x,t) dx 
 \nonumber\\
 &=&-\frac{\hbar^2}{m L}\sum_{n=1}^{\infty} \sum_{j=1}^{\infty}
 a_n^* a_j k_n k_j \beta_{n j} e^{i (\omega_n-\omega_j)t }.
\end{eqnarray}
We also easily have the result
\begin{eqnarray}
\label{Eq.3A-10}
&&\langle\Psi(t) \mid \hat{p} \mid \Psi(t)\rangle  \nonumber  \\
 &=&(- i \hbar)\ \frac{2}{ L}\ \sum_{n=1}^{\infty} \sum_{j=1, j\neq n}^{\infty}
 a_n^* a_j \frac{k_n k_j}{k_n^2-k_j^2} \beta_{n j} e^{i (\omega_n-\omega_j)t }.
\nonumber\\
\end{eqnarray}
It then follows that
\begin{eqnarray}
\label{Eq.3A-11}
  &&\frac{d}{dt}\langle\Psi(t)\mid \hat{p} \mid \Psi(t)\rangle 
  \nonumber\\
 &=& \frac{\hbar ^2}{ m L}\ \sum_{n=1}^{\infty} \sum_{j=1}^{\infty}
 a_n^* a_j k_n k_j \beta_{n j} e^{i (\omega_n-\omega_j)t }.
\end{eqnarray}
Comparing Eqs.~(\ref{Eq.3A-9}) and (\ref{Eq.3A-11}), we see that Ehrenfest's theorem is confirmed. We have thus directly verified Ehrenfest's theorem in the infinite square well.
Our results, Eqs.~(\ref{Eq.3A-8}), (\ref{Eq.3A-9}), are the same as those of  Rokhsar \citep{b.r9}.
%Our calculations are done by using directly the formula of $V(x) \Psi_n(x)$ in Eq.~(\ref{Eq.2A-6}). 
%We need not first do calculations above in the finite square well and then take the limit, $V_0 \to \infty$  \cite{b.r9}. 

For an example of Eq.~(\ref{Eq.3A-9}) , we choose $a_1= a_2 = 1/\sqrt{2}$, and $a_n=0$, otherwise. Then from Eq.~(\ref{Eq.3A-9}), we have
$\langle\Psi(t) \mid dV(\hat{x})/d\hat{x} \mid \Psi(t)\rangle =-(8 E_1/L)\cos(\omega_{12} t)$, where $\omega_{12}=(\omega_2-\omega_1) $.

%We note two of the special results. The first is that the expectation value 
%$ -\langle\Psi(t) \mid \frac{dV(\hat{x})}{d\hat{x}} \mid \Psi(t)\rangle$. The value is zero for any stationary state. 

%It had been argued that the infinite potential can not be simply described as the limit of a finite one \cite{b.r12}. 
%
%
%The potential energy $V(x)$ in a finite well is well defined. Simply taking the limit of the height of the barrier $V_0 \to \infty$ results the potential energy $V(x)$ defined in Eq.~(\ref{Eq.1A-1}), which does not have a well defined functional form. 
%It turns out that the limit should not be taken in the potential energy $V(x)$. Instead, the limit should be taken in the potential energy term $V(x) \Psi_n(x)$.

\section{\label{sec:4} More about the functional form of $V(x)$ }

In what follows, we try to obtain a functional form of $V(x)$ from Eq.~(\ref{Eq.2A-6}). We note that Eq.~(\ref{Eq.2A-6}) is not in a form that is symmetrical with respect to the two edges. We search for a more symmetrical one. We rewrite Eq.~(\ref{Eq.2A-6}) as
\begin{eqnarray}
&& V(x) \Psi_n (x)
= \frac{\hbar^2}{2m} ( A1+ A2).
\nonumber\\
&& A1= \sqrt {\frac{2}{L}} k_n  \delta(x).
\nonumber \\
&& A2= -\sqrt {\frac{2}{L}} k_n \cos(k_n L)  \delta(L-x).
\label{Eq.4A-1}
\end{eqnarray}
We can rearrange the two terms $A1$ and $A2$ expressed in terms of $u_n(x)$. 
For A1, using $k_n \delta(x) =[ \sin (k_n x)/x]  \delta(x) $, we then have $A1=[u_n(x)/x] \delta(x)$.
We also need to rewrite $A2$ in a similar way.
This can be done, as $u_n(x)=\sqrt {2/L} \sin(k_n  x)$ can also be written as

\begin{equation}
u_n(x)
= \sqrt \frac{2}{L} \sin[k_n  (x-L)] \cos (k_n L).
\label{Eq.4A-2}
\end{equation}
Above, we have used $\sin(k_n L)=0$ and $\cos(k_n L)= \pm 1$. 
Using
$k_n \delta(L-x) = [ \sin (k_n (L-x))/(L-x)]  \delta(L-x)$,
then we have: 
$ A2= \ [u_n(x)/(L-x)] \delta(L-x)$. 
Finally,  Eq.~(\ref{Eq.2A-6}) can be rewritten as 
\begin{equation}
V(x)\Psi_n(x) 
=  \frac{\hbar^2}{2m}  \left[\frac{\delta(x)}{x}+ \frac{\delta(L-x)}{(L-x)\ }\right] u_n(x).
\label{Eq.4A-3}
\end{equation}
Substituting $\Psi_n(x)$ in Eq.~(\ref{Eq.2A-1}) into Eq.~(\ref{Eq.4A-3}) then yields
\begin{equation}
V(x) \theta(x) \theta(L-x) u_n(x) 
=  \frac{\hbar^2}{2m}  \left[\frac{\delta(x)}{x}+ \frac{\delta(L-x)}{(L-x)\ }\right] u_n(x).
\label{Eq.4A-4}
\end{equation}
As the functions $u_n(x)$ form a complete set, Eq.~(\ref{Eq.4A-4}) can be extended to a general function $ \psi(x)$, which is composed from $u_n(x)$, that is    
\begin{eqnarray}
\psi(x)= \sum_{n=0}^{\infty} c_n u_n(x).
\label{Eq.4A-5}
\end{eqnarray}
Then we have
\begin{equation}
V(x) \theta(x) \theta(L-x)\psi(x) 
=  \frac{\hbar^2}{2m}  \left[\frac{\delta(x)}{x}+ \frac{\delta(L-x)}{(L-x)\ }\right] \psi(x).
\label{Eq.4A-6}
\end{equation}
This formula describes more precisely the property of the $V(x)$  in Eq.~(\ref{Eq.1A-1}).
%which contains the notation $\infty$. 
%The infinity in the  $V(x)$ is now with a precise specification.
Eq.~(\ref{Eq.4A-6}) together with Eq.~(\ref{Eq.4A-5}) are important results, as they provide a more precise starting point for the infinite square well model.

From Eq.~(\ref{Eq.4A-6}), we see that, naively, $V(x) \theta(x) \theta(L-x)$ behaves like having a form as $(\hbar^2/2m)  \left[\delta(x)/x + \delta(L-x)/(L-x) \right] $.
The factor $\theta(x) \theta(L-x)$ accompanied with $V(x)$ seems unavoidable.  
This factor is in fact needed, due to the divergence nature of the $V(x)$ outside the well. 
We may be curious on how to describe the degree of the infiniteness of the $V(x)$ in Eq.~(\ref{Eq.1A-1}).
Because  $V(x)$ is divergent outside the well, Eq.~(\ref{Eq.4A-6}) shows that this divergence is eliminated when multiplied by the function $\theta(x) \theta(L-x)$, as we see that the right side of  Eq.~(\ref{Eq.4A-6}) is well-defined.
Thus, the $\theta(x) \theta(L-x)$ term is needed to accompany with the $V(x)$ in order to have a regular result.
In other words, the $\theta(x) \theta(L-x)$ factor shows how to handle the infinity contained in the $V(x)$.
Eq.~(\ref{Eq.4A-6}) may then be viewed as a specification formula to specify the infinity in the  $V(x)$.

%It is then interesting to note that to define more clearly the $V(x)$ of an infinite square well, the $V(x)$ can not be defined alone, it must be accompanied with the factor $\theta(x) \theta(L-x)$. 

%
%

Eq.~(\ref{Eq.4A-6}) describes the property of the infinite barrier.
%Eq.~(\ref{Eq.4A-6}) is our most important result. This formula describes the more precise property of the notation $\infty$ used in Eq.~(\ref{Eq.1A-1}). 
This formula is obtained from the known eigenfunctions $\Psi_n(x)$, which are continuous at the two sides of the well. There are other types of solutions of the infinite square well, in which the continuity of $\Psi(x)$ at boundaries is not required \cite{yshuang}. We do not consider that case at present. 

\begin{acknowledgments}
The author is indebted to Prof. Young-Sea Huang for bringing in the attention and interests of this subject, and also for much help in discussions. The author would also like to thank Prof. Tsin-Fu Jiang and Prof. Wen-Chung Huang for much help. 
\end{acknowledgments}

\end{document}